\documentstyle{article}
\begin{document}
\def\today{}
\title{Higgs mass \\
and noncommutative geometry}
\author{Andrzej Sitarz \thanks{Partially supported by
KBN grant 2 P302 168 4} \thanks{E-mail: sitarz@if.uj.edu.pl} \\
Department of Field Theory \\
Institute of Physics \\
Jagiellonian University \\
Reymonta 4, 30-059 Krak\'ow, Poland}
\thispagestyle{empty}
\begin{titlepage}
\vfill
\maketitle
\begin{abstract}
We show that the description of the electroweak interactions
based on noncommutative geometry of a continuous and
a discrete space gives no special relations between the Higgs
mass and other parameters of the model. We prove that
there exists a gauge invariant term, linear in the curvature,
which is trivial in the standard differential geometry but
nontrivial in the case of the discrete geometry. The relations
could appear only if one neglects this term, otherwise one gets
the Lagrangian of the Standard model with the exact number of free
parameters.
\end{abstract}
\vfill
\end{titlepage}

\section{Introduction}

The noncommutative geometry has brought
a remarkable geometric picture explaining the
nature of the Higgs field and the symmetry
breaking mechanism \cite{CON1}-\cite{CGV} in
the Standard Model. The electroweak part of the
model is explained as originating from the product of the
continuous Minkowski space-time by a discrete
two-point space.

The Higgs field is a connection, which
arises from the geometry of the discrete
space. The latter corresponds to the left and
right fermions, with appropriate gauge groups associated
with each point. The bosonic part of the action comes
naturally as a Yang-Mills action, including the
quartic Higgs potential term.
The couplings of the Higgs field
to fermions arise from the usual couplings
between gauge potential and matter fields.

In this construction one recovers the
Lagrangian of the Standard Model with
one less free parameter. This has led
to speculations on the possible
theoretical predictions of the Higgs
mass, even though they were established
only on the classical level. However,
recent results \cite{QQQ} suggest that
they will not survive quantum corrections.

In this paper we shall advocate the idea that these
relations cannot be maintained already on the classical
level, due to the presence of a gauge-invariant term,
linear in the field strength. This term, together with
the Yang-Mills term, gives the original Lagrange function
of the Standard Model, with the same number of parameters.
Therefore, unless this term vanishes, there are no
classical relations between the Higgs mass and
other parameters of the Standard Model.

The existence of such term is characteristic
for the discrete geometry, therefore our arguments
do not change the relations obtained in some other
approaches \cite{DKM1}-\cite{BGW}.

\def\z{\hbox{\bf Z}}
\section{Gauge Theory on $M_4 \times \z_2$}

In this section we shall review the basic methods of constructing
the gauge theory on a product of the continuous Minkowski space
$M_4$ and the discrete two-point space $\z_2$. We shall present
here only the most substantial facts, more detailed account of
the construction can be found in \cite{CON1}-\cite{COQ3} and \cite{JA}.
We restrict ourselves only to the bosonic sector,
as it simplifies our considerations and does not influence
the results.

Let $x^\mu$ denote the coordinates on $M_4$ and let $y = \pm$
label the points of $\z_2$. The differentiation on the product
space $M_4 \times \z_2$ has the following form. If $f$ is a
function, then:
\begin{equation}
df = dx^\mu (\partial_\mu f) + \chi (\partial_+ f),
\end{equation}
where $dx^\mu$ are from the basis of one-forms on $M_4$ and $\chi$
is an object denoting the one-form on $\z_2$. The partial
derivative
$\partial_+$ is defined as follows:
\begin{equation}
(\partial_+ f)(x,y) \; = \; f(x,y) - f(x, -y),
\end{equation}

The remaining rules of the differential calculus define the
product of one forms:
\begin{eqnarray}
dx^\mu \otimes dx^\nu & = & - dx^\nu \otimes dx^\mu, \\
dx^\mu \otimes \chi & = & - \chi \otimes dx^\mu ,
\end{eqnarray}
the left multiplication by functions:
\begin{eqnarray}
f(x,y) \; dx^\mu & = & dx^\mu \; f(x,y), \\
f(x,y) \; \chi & = & \chi \; f(x, -y),
\end{eqnarray}
and the action of the external derivative $d$ on them:
\begin{eqnarray}
d (dx^\mu) & = & 0 \\
d (\chi) & = & 2 \; \chi \otimes \chi.
\end{eqnarray}

\def\a{\hbox{\bf A}}
The gauge potential $\a$ splits therefore into two parts,
the vector part $dx^\mu A^\mu$ and the scalar part
$\chi \Phi$. The names correspond to the behavior of the
coefficients $A_\mu$ and $\Phi$ under Lorentz transformations
in $M_4$. The unitarity of the gauge group enforces that
$A = - A^\star$, thus, since $dx^\mu$ are selfadjoint and
$\chi^\star = - \chi$, we obtain the following relations:
\begin{eqnarray}
A_\mu(x,y) & = & - A_\mu^\dagger(x,y), \\
\Phi(x,y) & = & \Phi^\dagger(x,-y). \label{r1}
\end{eqnarray}

\def\f{\hbox{\bf F}}
The curvature two-form $\f = d \a + \a \otimes \a$
splits into three terms:
\begin{equation}
\f \; = \; dx^\mu \otimes dx^\nu \; F_{\mu\nu} + dx^\mu \otimes \chi
\; {\cal F}_\mu + \chi \otimes \chi \; f,
\end{equation}
where $F_{\mu\nu}$ is the usual curvature tensor (obtained
independently for each value of $y$), ${\cal F}_\mu$ is a mixed
term, which includes the gradient of $\Phi$ and the coupling between
$A_\mu$ and $\Phi$. The last term, $f$, which depends only on $\Phi$
shall be of our primary interest, as it is responsible for the
appearance of the Higgs' quartic potential in the Yang-Mills action.

The precise form of $f$ is:
\begin{equation}
f(x,y) \; = \; \Phi(x,y) + \Phi(x,-y) + \Phi(x,-y) \Phi(x,y).
\label{r2}
\end{equation}

By introducing a shifted potential $\Psi(x,y) = 1+ \Phi(x,y)$ and
using relation (\ref{r1}), we may rewrite (\ref{r2}) as:
\begin{equation}
f(x,y) \; = \; \Psi^\dagger(x,y) \Psi(x,y) - 1.
\end{equation}

To construct the Yang-Mills action we need to have a metric $\eta$.
We take it to be a functional from the cartesian square of the
space of one-forms to the algebra of functions, which is
linear with respect to addition and left- and right-
multiplication by functions. In the considered case we
additionally postulate that the metric is diagonal, i.e.,
that $\eta(dx^\mu, \chi) = \eta(\chi, dx^\mu) = 0$.
We shall also assume here that
$$\eta^{\mu\nu}(x,y)  \; = \; \theta(y)
\tilde{\eta}^{\mu\nu}, $$
where $\tilde{\eta}$ is the standard Minkowski metric
and $\theta(y)$ is equal $\\gamma cos \theta_W$ for $y=+$
and $\gamma \sin \theta_W$ for $y=-$, $\theta_W$ is the Weinberg
angle and $\gamma$ an arbitrary positive number. The metric on the
discrete degrees of freedom is determined by one constant:
$$ \eta( \chi, \chi) = \alpha. $$
After taking such form of the metric the Yang-Mills Lagrange function
becomes:
\begin{equation} {\cal L}_{YM}(x,y)
\; = \; \gamma^2 \theta(y)^2 F_{\mu\nu} F^{\mu\nu} +
\alpha \gamma \theta(y)
{\cal F}^\dagger_\mu {\cal F}^\mu + \alpha^2 \left( \Psi^\dagger \Psi
- 1 \right)^2  \label{YM} \end{equation}
and the Lagrange function of the bosonic part of electroweak interactions
is obtained by taking the sum of the above for $y=+$ and $y=-$.

As can be seen directly in the formula (\ref{YM}) the quartic
potential of the Higgs field $\Psi$ has a very special form. Its shape is
determined only by one scaling constant, staying in front of the
whole term. Therefore one may establish a relation between the
mass of the field $\Psi$ and its vacuum expectation value,
which contributes to the masses of fields $A_\mu$.
Consequently, we should expect that the mass of the Higgs
field could be determined from the other parameters of Standard Model.

\section{Action term linear in curvature}

In this section we shall present an argument, which implies that
the suggested relations do not take place. Moreover, we shall show
that the approach based on noncommutative geometry gives us a precise
description of the Standard Model, with the same number of free
parameters.

Our idea is based on the observation that there
exists a possibility of adding an extra gauge invariant
term to the Yang-Mills action, which is linear in
the curvature $\f$. Let us proceed with the construction.

The curvature two-form $\f$ can be represented as a sum of the
tensor product of one forms:
\begin{equation}
\f = \sum_{i} u_i \otimes v_i.
\label{d1}
\end{equation}
Although this decomposition is not unique, we may observe that the
following quantity does not depend on its choice:
\begin{equation}
G(\f) \; = \; \sum_i \eta(u_i,v_i).
\end{equation}
This is due to the fact that both the tensor product and the metric
have the property of linearity. Moreover, this guarantees that
$G(\f)$ is gauge covariant and therefore its trace remains
gauge invariant. If we integrate it, we shall obtain an action,
which is linear in curvature:
\begin{equation}
S_l \; = \; \int \hbox{Tr} G(\f) + \hbox{c.c.} \label{ACT}
\end{equation} Of course, in the standard differential geometry such
action vanishes, because the decomposition (\ref{d1}) is
antisymmetrized and the metric is symmetric. However, in the case of
noncommutative geometry this no longer holds. In fact, in the model
considered in the previous section we have encountered the situation,
where the product of two one forms is not antisymmetric: $\chi \otimes
\chi \not= 0$. Therefore, in noncommutative geometry this action
(\ref{ACT}) could be nontrivial.

If we calculate the corresponding Lagrange function in our
particular case, the only nontrivial contribution comes
from the discrete degrees of freedom:
\begin{equation}
{\cal L}_l \; = \; 4 \alpha \left( \Psi^\star \Psi - 1 \right),
\label{BCT}
\end{equation}
where we have already integrated it over the discrete space $\z_2$.
Now, if we add this term to the Yang-Mills action, with an arbitrary
scaling parameter $\lambda$, the only change will be in the form
of the Higgs potential, which now can be rewritten as:
\begin{equation}
{\cal L}_{\hbox{Higgs}} \; = \; \beta \left(
\Psi^\star \Psi - \mu \right)^2,
\end{equation}
where $\mu = \frac{\lambda}{2\beta} -1$. The change of the constant
in front of the Higgs potential from $\alpha^2$ to $\beta$ reflects
that we may as well add the square of (\ref{BCT}), thus changing the
constant $\alpha^2$ to some other arbitrary value $\beta$.

We can see that this is precise form of the Higgs potential
with two free parameters, as usually assumed in the Standard Model.
Since we see no reason why these terms should be neglected, we may
conclude that they could contributes to the total action of the
model. Consequently, we obtain the full version of the Standard
Model, with no relations between the Higgs mass and other parameters.

\section{Conclusions}

As we have shown in the previous section, the existence of the
gauge invariant term linear in the curvature modifies the total
action of the Standard Model by increasing its number of free
parameters by one, back to the original value. This may be
disappointing, as the methods of noncommutative geometry has already
raised hopes that the number of free parameters might be reduced. In
fact, the indication that the Higgs mass is related to other constants
already on the Lagrangian level, was considered as an important result
of the noncommutative geometry approach. On the other hand, our
observation may be viewed as a completion of the derivation of the
precise form Standard Model of electroweak interaction from the
product of discrete and continuous geometry. It is remarkable
that one may obtain exactly the same classical Lagrangian from few
geometrical assumptions.

Since we have shown that the most general Lagrangian admits the
same number of parameters as the phenomenologically deduced one,
we may conclude that there is no specific formula for the Higgs mass.
Let us stress this fact that this hold only in case of the assumed
discrete geometry being responsible for appearance of Higgses. In
other approaches, which use the framework closer to the standard
differential geometry \cite{DKM1}-\cite{BGW}, the linear term
(\ref{ACT}) is trivial. Therefore some relations and estimates of the
Higgs mass are possible in such cases.

The geometrical description of the electroweak interactions and the
spontaneous symmetry breaking mechanism in terms of the product of
continuous and discrete geometry is an important step towards our
better understanding of these phenomena. Our result that one cannot
expect some basic relations between Higgs mass and other parameters
suggests that the original Lagrangian corresponds exactly to the
one derived by methods of noncommutative geometry. Of course, one
cannot exclude that the presented extra terms do not contribute
to the total action. This, however, could be verified only by
experiment, which would confirm the existence of the Higgs
particle and determine its mass.

\def\v#1{{\bf #1}}

\vfill
\end{document}